\documentclass[twocolumn,preprintnumbers,amsmath,amssymb]{revtex4}
\usepackage{graphicx}
\begin{document}
\title{Effect of ambient on the thermal parameters of micromachined bolometer}
\author{K. S. Nagapriya\footnote[1]{Electronic mail: ksnaga@physics.iisc.ernet.in}, A. K. Raychaudhuri\footnote[2]{Electronic mail: arup@physics.iisc.ernet.in, currently at S.N.Bose National Centre for Basic Sciences, Salt Lake, Kolkata 700098}}        
\affiliation{Department of Physics, Indian Institute of Science,  Bangalore 560 012,  India.}
\author{V. K. Jain, C. R. Jalwania and Vikram Kumar}
\affiliation{Solid State Physics Laboratory, Lucknow Road, New Delhi 100054, India}
\date{\today}
\begin{abstract}
The thermal characterization of bolometers is needed for optimal design as well as applications. In this paper, we present results of the effect of environment on the thermal properties of micromachined bolometers. We find that while in vacuum the thermal response can be represented by a single time constant, in presence of an ambient gas the thermal reponse can no longer be described by a single time constant. This will have a direct implication on frequency dependence of responsivity. We present a model to explain our data which involves the finite diffusion time in the ambient gas and the associated extra thermal mass.
\end{abstract}
\maketitle

Microbolometer arrays fabricated using Micro-Electro-Mechanical Systems (MEMS) technique ~\cite{Wood} are used for thermal image processing. Thermal response characterization of microbolometers is needed for optimal design.  Generally thermal characteristics are quantified through two measureable quantities - the thermal resistance of the bolometer  to the substrate  R$_{th}$ and the thermal relaxation time $\tau$. The thermal parameters are measured at a single bias voltage across the bolometer and also as a function of applied bias across the bolometer~\cite{Gu,Neuzil}. Application of excess bias leads to heating of the bolometer. This provides thermal characteristics at  elevated temperatures.    In this report we address the specific issue of  ambient. The ambient gas provides an additional thermal link of the bolometer to the substrate in addition to the support hinge and reduces  R$_{th}$. This changes the responsivity of the bolometer which is $\propto R_{th}$. By measuring the thermal characteristics in vacuum, air and Helium gas  ambient we have shown that the ambience not only changes the thermal link resistance R$_{th}$, but  more importantly,  changes the nature of  the time-temperature (t-T) response curve. In vacuum the (t-T) curve on application of a step heat input   shows exponential growth with one time constant while in air and He   it shows significant departure from such a simple behavior.  We could model the performance of the bolometer and have found that the  departure from a single time constant  arises because a finite mass of the gas ambient  gets associated with the thermal response of the bolometer giving rise to an extra heat capacity and an associated thermal relaxation time.  This changes the frequency response of the bolometer in presence of a cw sinusoidal signal input. 

The bolometer used in this investigation consists of an array of 4 $\times$ 4 elements, each of which is a Si$_3$N$_4$ membrane (50$\mu$m $\times$ 50$\mu$m)  grown on a Si wafer. A scanning electron microscope (SEM) image of a single element is shown in figure~1. The elements are connected to the main Si frame by Si$_3$N$_4$ hinges (5$\mu$m width x 2$\mu$m length x 1$\mu$m thickness). The thermal element is a Ti heater film 700$\AA$ thick and 4$\mu$m wide. 

\begin{figure}
\includegraphics[width=8cm]{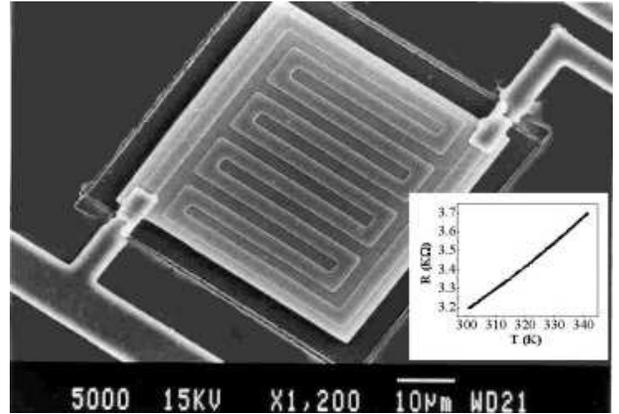}
\caption{\label{fig:figure1} SEM image of a single bolometer element. The inset shows the temperature dependence of the resistance of the Ti element.}
\end{figure}

  The experiments were carried out in a chamber that can be pumped down to a vacuum of  $10^{-5}$ torr. It can be filled with any desired gas upto a pressure of  1 atmosphere.  The thermal characteristics  were measured by  giving a step power input to the bolometer thermal element (Ti) and  recording the   temperature response of the bolometer as a function of time which we call the (t-T) curve.   For this we used the Ti element both as a heater and thermometer. To use the Ti element as the thermometer we calibrated  the  resistance of the Ti  as a function of temperature using a  standard Platinum thermometer  in the temperature range 100K to 350K. The TCR = 1/R(dR/dT) $\sim$ 4$\times$10$^{-3}$ /K.  The observed calibration is shown in the inset of figure~1.  The (t-T) curve  was obtained  by recording the voltage across the Ti element (which has a constant current bias)  using a 16 bit A/D card at a rate of 10K points/sec.  Step power input of amplitude P$_{in}$ leads to a rise in temperature $\Delta T$. In the steady state $\Delta$T = P$_{in}$ R$_{th}$. To analyze whether the (t-T) is governed by one time constant or not we have plotted the quantity $ln (1-(T(t)-T_{0})/ \Delta T)$ vs t where $T_{0}$ is the temperature of the bolometer in absence of the power input and is equal to the substrate temperature. When the (t-T) curve has a single time constant $(\tau)$ the $ln (1-(T(t)-T_{0})/ \Delta T)$ vs t curve is a straight line whose inverse slope gives $\tau$. Any departure from the single $\tau$ behavior will make this curve deviate from a straight line. 

\begin{figure}
\includegraphics[width=8cm]{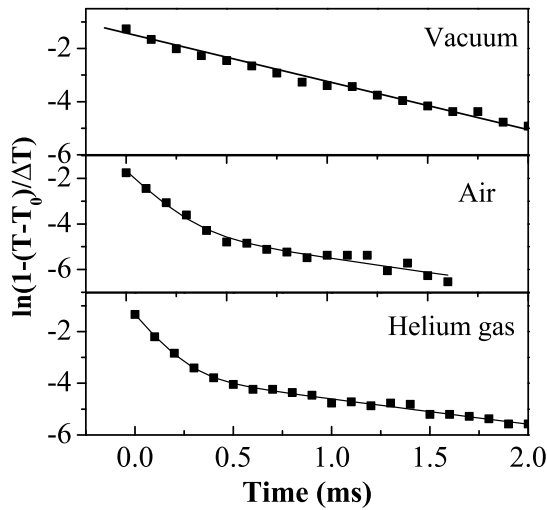}
\caption{\label{fig:figure2} Thermal response curve of the microbolometers in vacuum, air and He gas. Lines show fits from the model. Note the single relaxation time behavior of the response in vacuum and two time constants dominated behavior in an ambient gas.}
\end{figure}

Typical examples of observed $ln (1-(T(t)-T_{0})/ \Delta T)$ vs $t$ curves are shown in figure~2 for vacuum and with Air and Helium ambience. In vacuum the thermal response can be characterized by a single $\tau$, while in Air and He the curve  can be described by a combination of two time constants. From the given P$_{in}$ and the observed temperature rise $\Delta T$, we could obtain R$_{th}$ for all the experimental conditions- vacuum, Air and He ambience. The observed R$_{th}$ is shown in figure~3 as function of temperature.  In vacuum (pressure$<$10$^{-3}$ mbar) $R_{th} \approx  2.5 \times 10^{4} K/W$ at 300K and has a shallow temperature dependence.   In vacuum, in absence of other thermal pathways, R$_{th}$ will be determined by conduction through the hinges and the radiation contribution. Absence of a T$^{-3}$ type of a steep temperature dependence in R$_{th}$ shows that radiation makes a negligible contribution and R$_{th}$ will be  determined by the thermal conductivity $\Lambda$ of the hinge material. From the hinge dimension and the observed R$_{th}$ we calculate  $\Lambda \sim $ 30W/mK which is the same as the bulk value.  In air and He there is a large  reduction in   R$_{th}$  because the air  and He ambience provides additional thermal link. Understandably, He which has higher thermal conductivity shows a lower R$_{th}$. Thus in gas environment it is the gas  that determines the value of  R$_{th}$ .  

\begin{figure}
\includegraphics[width=8cm]{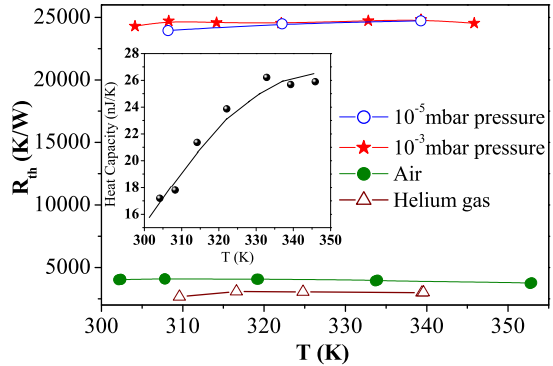}
\caption{\label{fig:figure3} The thermal resistance $R_{th}$ in vacuum, air and He gas as measured from the steady temperature rise $\Delta T$ and P$_{in}$. The inset shows the heat capacity of the membrane as determined from our experiment.}
\end{figure}

Next we analyze the observed thermal response curve shown in figure~2 using a simple model. The 
basic physics of the model is that  the presence of an ambient gas not only provides a thermal link (thus changing R$_{th}$), but also provides a thermal mass, the heat diffusion through which has a finite time which will lead to  additional time constants to the thermal response of the bolometer. The model is outlined below (see figure~4).  To simplify we have used the approach of "lumped circuit" model  where the thermal masses  are represented as capacitors and the thermal resistances are shown as electrical resistors. This is less rigorous than the actual  solution of  the heat diffusion equation  but it  is simple to solve and it captures the essential physics.  

\begin{figure}
\vspace{-3cm}
\includegraphics[width=10cm]{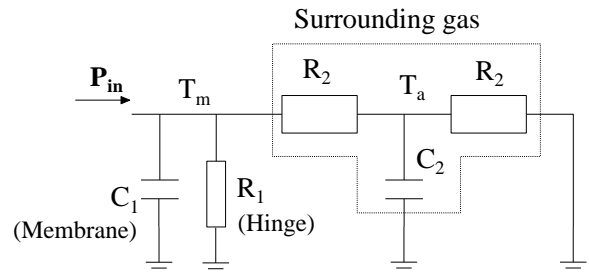}
\vspace{-3cm}
\caption{\label{fig:figure4} A thermal model of the bolometer in presence of an ambient.}
\end{figure}

In figure~4, $T_m$ (= $T(t)-T_{0}$)  is the temperature of the membrane with respect to bath (substrate) temperature $T_{0}$. $T_a$ is the average temperature of air with respect to the bath.  R$_1$ is the thermal resistance of the hinge and C$_1$ is the heat capacity of the membrane and  hinge combination. C$_2$ is the heat capacity of  ambient gas  that couples to the  bolometer and R$_2$ is the thermal link due to the gas. We have taken the thermal links of gas to membrane and  gas to  substrate both as $R_2$ for simplicity without sacrificing any feature.  The heat balance equations are 
\begin{equation}
P_{in}(t) = \frac{T_m - T_a}{R_2} + C_1\frac{dT_m}{dt} + \frac{T_m}{R_1}
\end{equation}

\begin{equation}
\frac{T_m - T_a}{R_2} = \frac{T_a}{R_2} + C_2\frac{dT_a}{dt}
\end{equation}

For a step heating  $P_{in}(t)$ = 0 for  t $<$ 0 and = P$_{in}$ for t $\geq $ 0.The solution to this set of coupled differential equations gives:
\begin{equation}
T_m = P_1 \{ 1 - exp(-gt)[ P_2 exp(-ft) + P_3 exp(ft)] \}
\end{equation}
where 

\hspace{0.1cm} 2g = ( (c$^{-1}$+ 2 )/$\tau _2$ ) + 1/$\tau _1$

\hspace{1cm} 4f$^2$ = (1/c$^2\tau _2^2$) + (2/c$\tau _1 \tau _2$) + (1/$\tau _1$ - 2/$\tau _2$)$^2$

\hspace{1cm} P$_1$ = 2P$_{in}$/ [C$_1\tau _2$(g$^2$-f$^2$)]

\hspace{1cm} P$_2$ = (1/2f) [ ($\tau _2$(g$^2$-f$^2$)/2) - (g-f) ]

\hspace{1cm} P$_3$ = (1/2f) [ (g+f) - $\tau _2$(g$^2$-f$^2$)/2) ]

Here we have defined $\tau _1$ = R$_1$C$_1$ and $\tau _2$ = R$_2$C$_2$ and $c = C_{1}/C_{2}$.The model gives two time constants, $(g-f)^{-1}$ and $(g+f)^{-1}$. In vacuum, the thermal resistance $R_{2}\rightarrow \infty$, R$_{th}$ = R$_{1}$ and one gets the thermal response being controlled by one time constant $\tau_{1}$. The fits to the data obtained using this model are shown in figure~2 as solid lines. In the case of the gas ambient the response curve with two time constants (eqn. 3) fits the observed data very nicely over the  complete range of $t$ over which data are taken. Typical $\tau_{1}$ at T=300K is $\approx$ 0.5 sec.   From the  observed value of  R$_{1}$ (obtained from the steady state value of $\Delta T$ and known value of P$_{in}$)  we could find the heat capacity C$_{1}$ of the bolometer as a function of T for the temperature range studied (figure~3 inset). C$_{1}$ has the  main contribution from the bulk of the heat capacity of the bolometer and approximately 1/3$^{rd}$ the heat capacity of the hinges. Using this we get a specific heat of $\approx$ 1.95 J/gmK for the material of the bolometer and the hinge. This is much larger than the value of specific heat of $\approx$ 1 J/gmK  for Si$_3$N$_4$. A value of $\sim$ 1.5J/gmK for the specific heat of Si$_3$N$_4$ has also been reported ~\cite{Eriksson}. 

At steady state $t \rightarrow \infty$ , $T_m = \Delta T = P_{1}$.  From the observed fits of the data to equation~3, the time constants $g^{-1}$, $f^{-1}$ and constants $P_2$ and $P_3$ can be found. Using the values of $\tau_1$, $R_1$ and $C_1$ as observed in the case of the vacuum we can find out $\tau_2$ , $R_2$ and $C_2$ , the parameters for the gas ambient. (In our model the parameters $\tau_1$, $R_1$ and $C_1$ depend only on the membrane and hinge material and are independent of the ambient gas). The parameters $\tau_2$, $R_2$ and $C_2$ are given in Table I.

\begin{table}
\caption{\label{tab:table1}Thermal parameters obtained from fits to the experimental data}
\begin{ruledtabular}
\begin{tabular}{ccccccc}
 Ambient&$\tau_1$&$R_1$&$C_1$&$\tau_2$&$R_2$&$C_2$\\
&$(msec)$&$(K/W)$&$(nJ/K)$&$(msec)$&$(K/W)$&$(nJ/K)$\\
\hline
Vacuum&0.53&$2.5\times 10^4$&20&$\infty$&$\infty$&-\\
Air&0.53&$2.5\times 10^4$&20&0.37&$4.0\times 10^3$&155\\
Helium&0.53&$2.5\times 10^4$&20&0.40&$3.1\times 10^3$&222\\
\end{tabular}
\end{ruledtabular}
\end{table}
The  heat capacity $C_2$  is due to the  extra thermal mass of gas around the bolometer through which the heat diffuses over the time scale of one to  few $\tau_2$. The volume of  the gas  calculated from the heat capacity is $\sim$ 1.34$\times$ 10$^{-10}$ m$^3$ for air and  $\sim$ 2.34$\times$ 10$^{-10}$ m$^3$ for He. If the volume in which the heat diffuses is taken to be a hemisphere (since the cavity below the membrane $\sim$ $2\mu m$), its  radius  ($r_g$) $\approx$ 400$\mu m$ for Air and 480$\mu m$ for He.  $r_g$ thus should be comparable to the thermal diffusion length $L_D$  of the gas.  The  $L_D$  is  estimated  from the relation $ L_{D} \approx \pi(\tau_{2} D)^{0.5}$ where $D$ is the diffusivity of  the gas.  From the standard $D = 5 \times 10^{-5} m^{2}sec^{-1}$ for air $D = 1.1 \times 10^{-4} m^{2}sec^{-1}$ for He we obtain $L_D \approx$ 440$\mu m$ for Air and $\approx$ 650$\mu m$ for He. It can be seen  for both the gases $r_{g} \sim L_{D}$. This is a good validation of the essential physics used  in the simple model. In a previous report  it was shown that the R$_{th}$  increases as the ambient is pumped and the vacuum is reached ~\cite{Eriksson}. However, the issue of two relaxation times has not been addressed to. It may be appreciated that the deviation from a single  relaxation time ($\tau$) dominated thermal response as observed in an ambience of gas will severely change the often used expression for the responsivity, $R (\omega) \propto (R_{th}/\sqrt{1+\omega^{2}\tau^{2}})$ which is strictly valid when the bolometer thermal response can be characterized by a single $\tau$. 

To summarize, in this paper we studied the effect of ambient gas on the thermal response of a microbolometer. We find that while in vacuum the thermal response can be characterized by a single time constant, the presence of ambient gas changes the nature of the thermal response and it needs two time constants to describe it. With a simple model we could connect the observed behavior to the physical parameters of the ambient gas.

\end{document}